\documentclass[prl,twocolumn,showpacs]{revtex4}
\usepackage{graphicx}
\usepackage{amsmath}
\usepackage{amsfonts}
\def\be{\begin{equation}}
\def\ba{\begin{eqnarray}}

\def\D{\Delta}

\def\o{\omega}

\def\i{\int}

\def\ee#1{\label{#1}\end{equation}}
\def\ea#1{\label{#1}\end{eqnarray}}

\begin{document}
\title{Complementary expressions for the entropy-from-work theorem}
\author{Michele Campisi}
\affiliation{Department of Physics,University of North Texas
Denton, TX 76203-1427, U.S.A.}
\date{\today}

\begin{abstract}
We establish an expression of the entropy-from-work theorem that
is complementary to the one originally proposed in [P. Talkner, P.
Hanggi and M. Morillo, arXiv:0707.2307]. In the original
expression the final energy is fixed whereas in the present
expression the initial energy is fixed.
\end{abstract}

\maketitle

In a recent and very interesting work \cite{talkner-2007}
\citeauthor{talkner-2007} have established the microcanonical
quantum fluctuation theorem. They proved that in quantum mechanics
one obtains the same fundamental equation that one finds
classically, namely:
\begin{equation}
\o_{E}(t_{0}) p_{t_{f},t_{0}}(E,w) = \o_{E+w}(t_{f})
p_{t_{0},t_{f}}(E+w,-w), \label{qC}
\end{equation}
where $\o_{E}(t_{0})$ and $\o_{E+w}(t_{f})$ are the densities of
states at energies $E$ and $E+w$ at times $t_0$ and $t_f$
respectively. The symbol $p_{t_{f},t_{0}}(E,w)$ denotes the
probability of moving from energy $E$ at time $t_0$ to energy
$E+w$ at time $t_f$, as a result of a protocol that changes the
Hamiltonian from $H(t_0)$ to $H(t_f)$. The symbol
$p_{t_{0},t_{f}}(E+w,-w)$ denotes the probability of starting from
$E+w$ at $t_f$ and ending up at $E$ at $t_0$ when the reversed
protocol is acted on the system.

Using Boltzmann's definition of entropy \footnote{For convenience
we set $k_{B}$, Boltzmann's constant equal to $1$.}: \be S(E,t) =
\ln\o_{E}(t), \ee{So} \citeauthor{talkner-2007} re-expressed Eq.
(\ref{qC}) as the microcanonical quantum version of Crook's
theorem \be \frac{p_{t_{f},t_{0}}(E,w)}{p_{t_{0},t_{f}}(E+w,-w)}=
e^{\left
    [S(E+w,t_{f})-S(E,t_{0}) \right ]}.
\ee{mcC}

Then, by expressing the final energy in terms of the initial
energy and work, and by integrating the exponentiated initial
entropy in Eq. (\ref{mcC}) over all possible values of work they
have been able to prove the following \emph{entropy-from-work}
theorem: \be e^{S(E_{f},t_{f}) } = N_{\rightarrow}(E_{f})  \langle
e^{S(E,t_{0}) }\rangle_{E_{f}}, \ee{SfS0}
 where
 \be N_{\rightarrow}(E_{f})
= \i dE
  \:p_{t_{f},t_{0}}(E,E_{f}-E)
\ee{N}

We label $N_{\rightarrow}$ with a right arrow to indicate that
$p_{t_{f},t_{0}}$ in the integrand refers to the forward protocol.

Note that the \emph{entropy-from-work} theorem of Eq. (\ref{SfS0})
is an equality that allows to extract an equilibrium property (the
final entropy) in terms of non-equilibrium work measurements. In
particular by running many experiments that end up with the same
energy $E_f$, and by measuring the work that has been performed in
each experiment, one is able to extract the value of final
entropy. Of course, since $w$ is a stochastic variable, the
initial energy of each experiment is not fixed and one has to
sample a certain range of initial energies.

In this communication we would like to point out another
non-equilibrium equality that follows from Eq. (\ref{mcC}). In
this equality the initial energy is fixed rather than the final,
and we look for the average exponentiated negative final entropy.
We have:
\begin{eqnarray}
\langle e^{-S(E+w,t_f)}\rangle_E &=& \int dw
p_{t_{f},t_{0}}(E,w)e^{-S(E+w,t_f)} \nonumber \\
   &=&  \int dw
p_{t_{f},t_{0}}(E,w) \o_{E+w}^{-1}(t_f) \nonumber \\
   &=& \int dw
p_{t_{0},t_{f}}(E+w,-w) \o_{E}^{-1}(t_0) \nonumber \\
   &=& e^{-S(E,t_0)}\int dE_f p_{t_{0},t_{f}}(E_f,E-E_f)
\end{eqnarray}
By defining
\begin{equation}\label{}
    N_{\leftarrow}(E) = \int d E_f
p_{t_{0},t_{f}}(E_f,E-E_f)
\end{equation}
we obtain:
\begin{equation}\label{eq:EFW2}
    \langle e^{-S(E+w,t_f)}\rangle_E =  N_{\leftarrow}(E) e^{-S(E,t_0)}
\end{equation}
This constitutes a second \emph{entropy-from-work} theorem by
means of which we can express the average exponentiated negative
final entropy in terms of non-equilibrium measurements of work.
The left arrow indicates that $p_{t_{0},t_{f}}$ in the integrand
refers to the backward protocol.

So we have established a second \emph{entropy-from-work} theorem
which is not opposed but rather complementary to the theorem
(\ref{SfS0}) provided in \cite{talkner-2007}. The two
complementary theorems can be put in the following symmetric form:
\begin{eqnarray}
\langle
e^{-\D S}\rangle_{E_{f}}&=& N_{\rightarrow}^{-1}(E_{f})\\
\langle e^{-\D S}\rangle_E &=& N_{\leftarrow}(E)
\end{eqnarray}
where $\Delta S = S(E_f,t_f)- S(E,t_{0})$. In the expression given
by \citeauthor{talkner-2007} the final energy $E_f$ is fixed and
the average is taken over all the processes that would end at that
energy. In the expression given here the initial energy $E$ is
fixed and the average is over all the processes that start from
that energy.

It is interesting to study the relations between the
\emph{entropy-from-work} theorem and the Second Law of
Thermodynamics. With reference to the form in Eq. (\ref{eq:EFW2}),
using Jensen equality one finds:
\begin{equation}\label{}
    \left\langle S(E+w,t_f)\right\rangle_E - S(E,t_0) \geq -\ln N_{\leftarrow}(E)
\end{equation}
Note that the previous equation does not ensure the positivity of
the entropy change. In fact, $N_{\leftarrow}(E)$ can be larger
than $1$, so that $-\ln N_{\leftarrow}(E)$ can be negative. This
means that the final expectation of Boltzmann entropy in Eq.
(\ref{So}) can be lower than the initial value. One example where
this happens is the 1D harmonic oscillator with changing
frequency. For this system the density of states does not depend
on energy and is proportional to the inverse of the frequency.
Thus for any protocol of increasing frequency the change in
Boltzmann entropy is negative. Recent works
\cite{Campisi08bis,Sasa00} suggest that the final expectation of
microcanonical entropy could be proved to be always larger than
the initial value if the alternative definition where the density
of states is replaced by the volume of phase space is employed.
This has been already proved in general for the canonical initial
condition \cite{Campisi08,Campisi08bis} and in high dimensional
chaos for the microcanonical initial condition \cite{Sasa00}.
Nonetheless no associated entropy-from-work theorem has been
reported yet that employs this alternative definition of entropy.

With reference to \citeauthor{talkner-2007} form (\ref{SfS0}), in
an analogous way one also finds:
\begin{equation}\label{}
     S(E_f,t_f) - \left\langle S(E,t_0)\right\rangle_{E_f} \geq \ln N_{\rightarrow}(E_f)
\end{equation}
which again does not put any constraint to the positivity of
averaged entropy change.

To summarize, we have established another form of the
entropy-from-work theorem that refers to the case of fixed initial
energy. This is complementary to the expression given by
\citeauthor{talkner-2007} where instead the final energy is fixed.
The relations between the entropy-from-work theorems and the
second law of thermodynamics have been discussed too.


\end{document}